\documentclass[12pt]{article}
\usepackage{amsmath,amssymb}
\usepackage{graphicx}
\allowdisplaybreaks[1]

\makeatletter
 
  \@addtoreset{equation}{section}
 \makeatother
\newcommand{\cN}{{\cal N}}
\newcommand{\cL}{{\cal L}}
\newcommand{\cO}{{\cal O}}

\newcommand{\Z}{\mathbb{Z}}

\newcommand{\nn}{\nonumber}

\newcommand{\Tr}{{\rm Tr}\,}

\newcommand{\del}{\partial}

\newcommand{\diag}{{\rm diag}\,}

\allowdisplaybreaks
\begin{document}
\begin{titlepage}

\setcounter{page}{0}
\renewcommand{\thefootnote}{\fnsymbol{footnote}}

\begin{flushright}
OIQP-14-03
\end{flushright} 
\vspace*{5mm}

\begin{center}
{\large\bf
Lattice Formulation for 2d $\cN=(2,2), \,(4,4)$ Super Yang-Mills Theories
without Admissibility Conditions
}

\vspace{10mm}
So Matsuura$^{1}$%
\footnote{\tt s.matsu@phys-h.keio.ac.jp}
and Fumihiko Sugino ${}^{2}$
\footnote{\tt fumihiko\_sugino@pref.okayama.lg.jp} 
\vspace{10mm}

{\em 
$^{1}$ Department of Physics, 
and Research and Education Center for Natural Science, 
Keio University, 4-1-1 Hiyoshi, Yokohama, 223-8521, Japan\\
$^{2}$ Okayama Institute for Quantum Physics, 
Kyoyama 1-9-1, Kita-ku, Okayama 700-0015, Japan}
\end{center}

\vspace{20mm}
\centerline{{\bf Abstract}}
\vspace{3mm}
{\small 
We present a lattice formulation 
for two-dimensional ${\cal N}=(2,2)$ and $(4,4)$ supersymmetric Yang-Mills theories 
that resolves vacuum degeneracy for gauge fields 
without imposing admissibility conditions. 
Cases of $U(N)$ and $SU(N)$ gauge groups are considered, 
gauge fields are expressed by unitary link variables, 
and one or two supercharges are preserved on the two-dimensional square lattice.  
There does not appear fermion doubler, and no fine-tuning is required 
to obtain the desired continuum theories in a perturbative argument. 
This formulation is expected to serve as a more convenient basis for numerical simulations. 
The same approach will also be useful to other two-dimensional 
supersymmetric lattice gauge theories 
with unitary link variables constructed so far -- for example, 
$\cN=(8,8)$ supersymmetric Yang-Mills theory and $\cN=(2,2)$ supersymmetric QCD. 
}
\end{titlepage}
\newpage

\setcounter{footnote}{0}

\section{Introduction}
Lattice formulation is one of the most powerful ways to investigate 
nonperturbative aspects of quantum field theories, in particular gauge field theories. 
It gives not only rigorous definitions of quantum field theories, 
but also yields an environment to carry out numerical ``experiments'' on the theories. 
This approach has achieved a great success for lattice QCD. 
As a step along this line,  
it is quite natural and significant to head for lattice regularizations of 
supersymmetric gauge theories. 
In spite of difficulties of realizing supersymmetry on lattice, 
several lattice formulations of supersymmetric gauge theories 
which need no fine-tuning in taking the continuum limit have been developed. 
In particular, for theories with extended supersymmetries, 
some of supercharges are exactly preserved on lattice by applying the so-called orbifolding procedure
~\cite{Kaplan:2002wv,Cohen:2003xe,Kaplan:2005ta,Endres:2006ic,Giedt:2006dd,Matsuura:2008cfa,Joseph:2013jya} 
or the topological twists
~\cite{Catterall:2003wd,Sugino:2003yb,Sugino:2004qd,Sugino:2004uv,Sugino:2006uf,Sugino:2008yp,Kikukawa:2008xw,Kadoh:2009yf}  
to the discretization. 

For ${\cal N}=1$ pure supersymmetric Yang-Mills (SYM) theories 
in three and four dimensions, 
preserving chiral symmetry rather than supersymmetry on the lattice 
plays a key role to restore the supersymmetry and all the other symmetries 
in the continuum limit~\cite{Maru:1997kh,Giedt:2009yd}. 
Except for these cases, however, 
it will be hard to consider lattice regularizations 
for three- and four-dimensional supersymmetric theories 
in such a way that the continuum limit requires no tuning.  
In fact, 
the number of symmetries preserved on the lattice 
is generally too small to forbid relevant operators that prevent the lattice theory from restoring 
all the symmetries (including supersymmetries) of the target theory in the continuum limit. 
As an approach to circumvent this issue, 
a hybrid regularization  has been proposed for four-dimensional $\cN=2, \,4$ SYM 
theories~\cite{Hanada:2010kt,Hanada:2010gs,Hanada:2011qx}, where 
two different discretizations by lattice and matrix~\cite{Berenstein:2002jq,Das:2003yq,Myers:1999ps} 
are combined. 
Regarding to the planar limit, four-dimensional $\cN=4$ SYM theory 
can be obtained by using a large-$N$ reduction technique~\cite{Ishii:2008ib,Ishiki:2011ct}. 
%

In 
\cite{Kaplan:2002wv,Cohen:2003xe,Kaplan:2005ta,Endres:2006ic,Giedt:2006dd,Matsuura:2008cfa,Joseph:2013jya}, 
various theories 
have been formulated on exotic lattices by the 
orbifolding procedure from SYM matrix theories. 
In 
\cite{Catterall:2003wd,DAdda:2004jb,Nagata:2008zz}, 
essentially the same formulations as the orbifold lattice theories 
have been independently developed 
\cite{Unsal:2006qp,Catterall:2007kn,Damgaard:2007eh}
from different approaches.%
~\footnote{
For a review, see \cite{Catterall:2009it}. 
} 
In these formulations, the bosonic link variables are not unitary but complex matrices 
and the lattice spacing is introduced 
by fixing the trace part of these variables at a specific point in the flat directions. 
As a result, gauge groups which the formulations allow are $U(N)$ rather than $SU(N)$. 
Differently from the continuum theory, 
the overall $U(1)$ modes in $U(N)$ are always coupled with the remaining $SU(N)$ modes in the lattice theory. 
In particular, there are zero-modes in the $U(1)$ sector of fermions.  
In numerical simulations, therefore, we must introduce a large mass 
in the $U(1)$ part of the complex link variables in order to fix the 
lattice spacing and take care of the fermionic zero-modes in computing 
the Dirac matrix~\cite{Kanamori:2008bk,Hanada:2009hq,Catterall:2010fx}. 
It is important to check decoupling of the $U(1)$ sector 
in the continuum limit in order to confirm that the correct continuum theory is obtained.

In 
\cite{Sugino:2003yb,Sugino:2004qd,Sugino:2004uv,Sugino:2006uf,Sugino:2008yp,Kikukawa:2008xw}, 
one of the authors of the present paper (F.S.) discretized topologically twisted 
gauge theories with preserving one or two supercharges. 
One characteristic feature of this formulation is that 
lattice gauge fields are expressed by compact link variables on the hypercubic lattice 
as in the conventional lattice gauge theories, 
which will be more convenient for numerical 
simulations~\cite{Suzuki:2007jt,Kanamori:2007ye,Kanamori:2007yx}. 
In addition, it is valid for both of the gauge groups $U(N)$ and $SU(N)$. 
In particular, for the gauge group $SU(N)$, we do not need to be bothered about the $U(1)$ sector 
mentioned above. 
On the other hand, 
we have to take care of vacuum degeneracy of lattice gauge fields. 
In \cite{Sugino:2004qd}, an admissibility condition is imposed 
in order to single out the physical vacuum from the other unphysical vacua. 
Although this prescription works well, 
it yields complicated simulation codes in practice. 

In this paper, we modify the formulation 
for two-dimensional $\cN=(2,2)$ and $(4,4)$ SYM theories~\cite{Sugino:2003yb,Sugino:2004qd} 
so that the vacuum degeneracy is resolved without imposing the admissibility condition. 
It is possible with keeping relevant symmetries to ensure no fine-tuning in the continuum limit 
for both of the gauge groups $U(N)$ and $SU(N)$. 
The modified formulation simplifies lattice actions, which will serve a more convenient basis 
for numerical simulations.  

This paper is organized as follows. 
In the next section, we give a short review of the lattice formulation for 
two-dimensional $\cN=(2,2)$ and $(4,4)$ SYM theories in~\cite{Sugino:2003yb,Sugino:2004qd}. 
In section~\ref{sec:N=2_woAD}, the modification of the $\cN=(2,2)$ theory is discussed for the gauge 
groups $U(N)$ and $SU(N)$. Convenient expressions for actual numerical simulations are also presented. 
In section~\ref{sec:N=4_woAD}, the $\cN=(4,4)$ theory is modified in parallel with the $\cN=(2,2)$ case. 
The results obtained so far are summarized and some future directions are discussed in 
section~\ref{sec:summary}. 
Appendix~\ref{app:wall} is devoted to a proof that potential barriers of the $SU(N)$ lattice gauge action 
infinitely grow away from the physical vacuum, which makes any unphysical vacuum ineffective.


\section{Brief review of the lattice formulation for 2d $\cN=(2,2)$ and $(4,4)$ SYM theories}
\label{sec:review}
In this section, we present a brief review of 
the lattice formulation for two-dimensional $\cN=(2,2)$ and $(4,4)$ SYM theories 
constructed in~\cite{Sugino:2003yb,Sugino:2004qd}. 
In what follows, $x$ denotes a site of the two-dimensional square lattice $\Z_L^2$, 
where $L$ is the number of the sites in one direction. 
Lattice gauge fields are expressed by the group-valued variables $U_\mu(x)=e^{iaA_\mu(x)}$ on the link 
$(x, x+\hat{\mu})$, 
while all the other fields are algebra-valued variables put on the sites.  

\subsection{2d $\cN=(2,2)$ SYM theory on the lattice}
Field contents of $\cN=(2,2)$ SYM theory on the two-dimensional lattice~\cite{Sugino:2003yb,Sugino:2004qd} are as follows.  
Bosonic variables are 
the unitary link variables $U_\mu(x)$, 
scalar variables%
~\footnote{$\phi(x)$ and $\bar{\phi}(x)$ can be treated as independent hermitian matrices in path-integrals 
of the theory.} 
$\phi(x)$ and $\bar\phi(x)$ 
and a hermitian auxiliary field $H(x)$. 
Fermionic variables are denoted by $\{ \psi_\mu(x), \chi(x), \eta(x) \}$. 
Note that, when the gauge group is $SU(N)$, all the fields excluding $U_\mu(x)$ are 
traceless. 
These lattice variables are connected by the supersymmetry transformation:  
\begin{align}
QU_\mu(x) &= i\psi_\mu(x)U_\mu(x), \quad 
Q\psi_\mu(x) = iD_\mu\phi(x) + i \psi_\mu(x) \psi_\mu(x), \nn \\
Q\bar\phi(x) &= \eta(x), \quad Q\eta(x) = [ \phi(x), \bar\phi(x) ], \nn \\
Q\chi(x) &= H(x), \quad QH(x) =  [ \phi(x), \chi(x) ], \quad Q\phi(x)=0, 
\label{Q_SUSY}
\end{align}
where $D_\mu$ represents the covariant forward difference given by
$D_\mu \varphi(x) \equiv U_\mu(x)\varphi(x+\hat{\mu})U_\mu(x)^\dagger -\varphi(x)$ 
for any adjoint field $\varphi(x)$ on the site. 
Note that $Q$ is nilpotent up to infinitesimal gauge transformation with the parameter $\phi(x)$. 
The lattice action can be expressed as a $Q$-exact form: 
\begin{align}
S_{\rm lat}^{(2,2)} & = Q \frac{1}{2g_0^2}\sum_x \Tr \Bigl[
\frac{1}{4} \eta(x) [\phi(x), \bar\phi(x)] - i\chi(x)\left( \Phi(x) + i H(x) \right)
-i \sum_{\mu=1}^2 \psi_\mu(x) D_\mu\bar\phi(x)
\Bigr] \nn\\
 & = \frac{1}{2g_0^2}\sum_x \Tr \Bigl[\frac14[\phi(x), \bar{\phi}(x)]^2 +H(x)^2 -iH(x) \Phi(x) 
+\sum_{\mu=1}^2 D_\mu\phi(x) D_\mu\bar{\phi}(x) \nn \\
 & \hspace{26mm} -\frac14\eta(x)[\phi(x), \eta(x)]-\chi(x)[\phi(x), \chi(x)] \nn \\
 & \hspace{26mm} -\sum_{\mu=1}^2 \psi_\mu(x)\psi_\mu(x)\left(\bar{\phi}(x)
+U_\mu(x)\bar{\phi}(x+\hat{\mu})U_\mu(x)^\dagger\right) \nn \\
 & \hspace{26mm} +i\chi(x)\,Q\Phi(x) +i\sum_{\mu=1}^2 \psi_\mu(x)D_\mu\eta(x)\Bigr], 
\label{Slat_N=2} 
\end{align}
where $\Phi(x)$ is a hermitian matrix depending on the plaquette variables, 
\begin{equation}
U_{12}(x)\equiv U_1(x) U_2(x+\hat{1}) U_1(x+\hat{2})^\dagger U_2(x)^\dagger 
\quad \mbox{and}\quad  U_{21}(x)=U_{12}(x)^\dagger. 
\end{equation}

After the auxiliary field is integrated out, 
the action of the gauge fields is given by
\begin{eqnarray}
S_G^{U(N)} & = & \frac{1}{8g_0^2} \sum_{x} \Tr \left( \Phi(x)^2 \right) \quad 
\mbox{for} \quad U(N), \label{SG_UN}\\
S_G^{SU(N)} & = & \frac{1}{8g_0^2} \sum_{x} \Tr \left( \left\{\Phi(x)-\left(\frac{1}{N}\Tr \Phi(x)\right) {\bf 1}_N\right\}^2 \right)  \quad
\mbox{for} \quad SU(N). 
\label{SG_SUN}
\end{eqnarray}
For the gauge group $SU(N)$, since the auxiliary field $H(x)$ is traceless, 
only the traceless part of $\Phi(x)$ contributes to the action. 
In order to obtain the correct continuum action, the expansion 
by the lattice spacing $a$ of $\Phi(x)$ around $U_{12}(x)={\bf 1}_N$ must be 
\begin{equation}
\Phi(x) = 2a^2 F_{12}(x) + {\cal O}(a^3), 
\label{expansion}
\end{equation}
where $F_{12}(x) \equiv \del_1 A_2(x)-\del_2 A_1(x)+i[A_1(x), A_2(x)]$. 
The simplest choice satisfying (\ref{expansion}) is  
\begin{equation}
 \Phi(x) = -i \left( U_{12}(x) - U_{21}(x) \right). 
 \label{naive phi}
\end{equation}
However, this causes a problem: 
there are a number of unphysical degenerate vacua~\cite{Sugino:2003yb}. 
In the diagonal gauge, 
\begin{equation}
U_{12}(x)=\diag(e^{i\theta_1(x)},\cdots,e^{i\theta_N(x)}), 
\label{diagonal}
\end{equation}
(\ref{naive phi}) becomes 
$\Phi(x)=\diag( 2\sin \theta_1(x),\cdots, 2\sin \theta_N(x))$
which has zeros at $\theta_i(x)=0$ and $\pi$. 
There remain $(N+1)^{L^2}$ degenerate vacua for the gauge group $U(N)$ 
after a subgroup of the gauge symmetry (permutations of the eigenvalues) is taken into account. 
See \cite{Sugino:2004qd,Kanamori:2012et} for vacuum degeneracy in the case of $SU(N)$. 
In \cite{Sugino:2004qd}, an admissibility 
condition is introduced in order to single out the trivial vacuum $U_{12}={\bf 1}_N$ as follows. 
When $||1-U_{12}(x)||<\epsilon$ for ${}^\forall x$, 
$S_{\rm lat}^{(2,2)}$ is defined by (\ref{Slat_N=2}) with the choice of $\Phi(x)$ as  
\begin{equation}
 \Phi(x)=\frac{  -i \left( U_{12}(x) - U_{21}(x) \right) }
 {1-\frac{1}{\epsilon^2}|| 1- U_{12}(x) ||^2 },
 \label{admissibility}
\end{equation}
otherwise $S_{\rm lat}^{(2,2)}=+\infty$. 
Here, $|| \cdot ||$ is a norm of a matrix defined by 
$||A||\equiv \sqrt{\Tr(AA^\dagger)}$, and $\epsilon$ is a positive number 
chosen in the range $0<\epsilon<2$ for $U(N)$, $0<\epsilon<2\sqrt{2}$ for $SU(N)$
with $N=2,3,4$ and $0<\epsilon<2\sqrt{N}\sin\left(\frac{\pi}{N}\right)$ for $SU(N)$
with $N\ge5$. 
Thanks to this admissibility condition, we can restrict the value of the plaquette 
variables to the range $|| 1- U_{12}(x) || < \epsilon$ without breaking the $Q$ supersymmetry, 
and as a result we can single out the trivial vacuum 
$U_{12}={\bf 1}_N$ from the other unphysical vacua.

\subsection{2d $\cN=(4,4)$ SYM theory on the lattice}
In the lattice model for two-dimensional $\cN=(4,4)$ SYM theory~\cite{Sugino:2003yb,Sugino:2004qd}, 
there are scalar fields $B(x)$, $C(x)$, $\phi_\pm(x)$,%
~\footnote{Here, we write the scalar fields $\phi(x)$ and $\bar{\phi}(x)$ 
in~\cite{Sugino:2003yb,Sugino:2004qd} as 
$\phi_+(x)$ and $\phi_-(x)$, respectively.} 
and auxiliary fields $\tilde{H}_\mu(x)$ and $H(x)$ 
in addition to the link variables $U_\mu(x)$. 
Fermionic fields are $\psi_{\pm\mu}(x)$, $\chi_\pm(x)$ and $\eta_\pm(x)$. 
The supersymmetry transformations preserved on the lattice 
\begin{eqnarray}
& & Q_\pm U_\mu(x) = i\psi_{\pm\mu}(x) U_\mu(x), \qquad 
Q_\pm \psi_{\pm \mu}(x) = i\psi_{\pm\mu}(x)\psi_{\pm\mu}(x) \pm iD_\mu\phi_\pm (x), \nn \\
& & Q_\mp \psi_{\pm\mu}(x) = \frac{i}{2}\{\psi_{+\mu}(x), \psi_{-\mu}(x)\} +\frac{i}{2}D_\mu C(x) 
\mp \tilde{H}_\mu(x), \nn \\
& & Q_\pm \tilde{H}_\mu(x) = -\frac12\left[\psi_{\mp\mu}(x), \phi_\pm(x)
+U_\mu(x)\phi_\pm(x+\hat{\mu})U_\mu(x)^\dagger\right] \nn \\
& & \hspace{23mm} \pm\frac14\left[\psi_{\pm\mu}(x), C(x)+U_\mu(x)C(x+\hat{\mu})U_\mu(x)^\dagger\right] \nn \\
& & \hspace{23mm} \mp\frac{i}{2}D_\mu \eta_\pm(x) +\frac{i}{2}\left[\psi_{\pm\mu}(x), \tilde{H}_\mu(x)\right]
\pm\frac14\left[\psi_{\pm\mu}(x)\psi_{\pm\mu}(x), \psi_{\mp\mu}(x)\right], 
\nn \\
 & & Q_\pm B(x)=\chi_\pm(x), \quad Q_\pm \chi_\pm(x) = \pm [\phi_\pm(x), B(x)], \quad 
Q_\mp \chi_\pm(x)= \frac12[C(x), B(x)]\mp H(x), \nn \\
 & & Q_\pm H(x) = [\phi_\pm(x), \chi_\mp(x)] \pm \frac12[B(x), \eta_\pm(x)] \mp \frac12[C(x), \chi_\pm(x)], 
\nn \\
 & & Q_\pm C(x)=\eta_\pm(x), \quad Q_\pm\eta_\pm(x) = \pm [\phi_\pm(x), C(x)], \quad 
Q_\mp \eta_\pm(x) = \mp [\phi_+(x), \phi_-(x)], \nn \\
 & & Q_\pm\phi_\pm(x)=0, \quad Q_\mp \phi_\pm(x) = \mp\eta_\pm(x) 
\label{Qpm_SUSY} 
\end{eqnarray}
are nilpotent in the sense that 
\begin{eqnarray}
Q_+^2 &= & (\mbox{infinitesimal gauge transformation with the parameter $\phi_+(x)$}), \nn \\
Q_-^2 &= & (\mbox{infinitesimal gauge transformation with the parameter $-\phi_-(x)$}), \nn \\
\{Q_+, Q_-\} &= & (\mbox{infinitesimal gauge transformation with the parameter $C(x)$}). 
\end{eqnarray}
The lattice action can be expressed as the $Q_+$ and $Q_-$ transformations of gauge invariant terms:  
\begin{equation}
S_{\rm lat}^{(4,4)}= Q_+Q_-\frac{1}{2g_0^2}\sum_x\Tr\Bigl[-iB(x)\Phi(x)
-\sum_{\mu=1}^2 \psi_{+\mu}(x)\psi_{-\mu}(x)-\chi_+(x)\chi_-(x)-\frac14\eta_+(x)\eta_-(x)\Bigr], 
\label{Slat_N=4}
\end{equation}
which is written down more explicitly as 
\begin{align}
S_{\rm lat}^{(4,4)}= &\frac{1}{2g_0^2}\sum_x\Tr\Bigl[-i\left(\frac12[C(x), B(x)]+H(x)\right)\Phi(x) +H(x)^2 
\nn \\
 & \hspace{12mm}+i\chi_-(x)Q_+\Phi(x)-i\chi_+(x)Q_-\Phi(x)-iB(x)Q_+Q_-\Phi(x) \nn \\
 & \hspace{12mm}-[\phi_+(x), B(x)][\phi_-(x),B(x)]-\frac14[C(x), B(x)]^2 \nn \\
 & \hspace{12mm}+\chi_+(x)[\phi_-(x), \chi_+(x)]-\chi_-(x)[\phi_+(x), \chi_-(x)]+\chi_+(x)[C(x), \chi_-(x)] 
\nn \\
 & \hspace{12mm}-\chi_-(x)[B(x), \eta_+(x)] -\chi_+(x)[B(x),\eta_-(x)] \nn \\
 & \hspace{12mm}+\frac14 [\phi_+(x), \phi_-(x)]^2 -\frac14[\phi_+(x), C(x)][\phi_-(x), C(x)] \nn \\
 & \hspace{12mm}-\frac14 \eta_-(x) [\phi_+(x), \eta_-(x)]+\frac14\eta_+(x)[\phi_-(x), \eta_+(x)] 
 -\frac14 \eta_+(x)[C(x), \eta_-(x)]\Bigr] \nn \\
 &+\frac{1}{2g_0^2}\sum_x \sum_{\mu=1}^2\Tr\Bigl[\tilde{H}_\mu(x)^2
+\frac12\psi_{+\mu}(x)\psi_{+\mu}(x)\psi_{-\mu}(x)\psi_{-\mu}(x) +D_\mu\phi_+(x)D_\mu\phi_-(x) \nn \\
 &\hspace{24mm}+\frac14\left(D_\mu C(x)\right)^2 +i\psi_{+\mu}(x)D_\mu\eta_-(x) 
 +i\psi_{-\mu}(x)D_\mu\eta_+(x) \nn \\
 &\hspace{24mm} -\psi_{+\mu}(x)\psi_{+\mu}(x)\left(\phi_-(x)+U_\mu(x)\phi_-(x+\hat{\mu})U_\mu(x)^\dagger\right) \nn \\
 & \hspace{24mm}+\psi_{-\mu}(x)\psi_{-\mu}(x)\left(\phi_+(x)+U_\mu(x)\phi_+(x+\hat{\mu})U_\mu(x)^\dagger\right) \nn \\
 & \hspace{24mm}-\frac12 \{\psi_{+\mu}(x), \psi_{-\mu}(x)\} \left(C(x)+U_\mu(x)C(x+\hat{\mu})U_\mu(x)^\dagger\right)
 \Bigr]. 
\end{align}
Here, $\Phi(x)$ has the same property as that in the $\cN=(2,2)$ case, 
which provides the gauge field action 
(\ref{SG_UN}) or (\ref{SG_SUN}) after $H(x)$ is integrated out. 
The problem of vacuum degeneracy mentioned in the $\cN=(2,2)$ case commonly arises in the $\cN=(4,4)$ case. 
Thus, imposing the same admissibility condition as in (\ref{admissibility}) resolves 
the problem~\cite{Sugino:2004qd}.

\section{2d $\cN=(2,2)$ lattice SYM theory without the admissibility condition }
\label{sec:N=2_woAD}
As mentioned in the introduction, it is better to find an expression of $\Phi(x)$ 
that has a unique vacuum at $U_{12}={\bf 1}_N$ without imposing the admissibility condition. 
At the first sight, however, it seems impossible 
because of the condition (\ref{expansion}) and the periodicity of the plaquette variable. 
In fact, in the diagonal gauge (\ref{diagonal}), we can express $\Phi(x)$ as 
$\Phi(x)=\diag( f(\theta_1(x)),\cdots, f(\theta_N(x))  )$, where $f(\theta)$
is a periodic function with the period $2\pi$. 
The condition (\ref{expansion}) means that $f(\theta)$ is linear around the origin: 
$f(\theta)=2\theta+{\cal O}(\theta^2)$, 
and the periodicity of the plaquette variable suggests $f(-\pi)=f(\pi)$. 
There must be at least one zero in $\theta\in (-\pi,\pi]$ except for the origin 
as long as $f(\theta)$ is a regular function. 
However, if the regularity of $f(\theta)$ is relaxed, 
one can construct a desirable expression of $\Phi(x)$. 
In this section, we provide such an expression in two-dimensional $\cN=(2,2)$ lattice SYM theory 
for both of the gauge groups $U(N)$ and $SU(N)$. 

\subsection{$U(N)$ gauge group}

For the gauge group $U(N)$, 
we show that the expression 
\begin{align}
\Phi^{(1)}(x) 
&=2i \left( 2 - U_{12}(x) - U_{21}(x) \right) \left( U_{12}(x) - U_{21}(x) \right)^{-1} \nn \\
&\quad +2i \left( U_{12}(x) - U_{21}(x) \right)^{-1} \left( 2 - U_{12}(x) - U_{21}(x) \right) \nn \\
&\equiv \frac{ 4 \left( 2 - U_{12}(x) - U_{21}(x) \right) }
{ -i \left( U_{12}(x) - U_{21}(x) \right) } 
\label{UNphi}
\end{align}
has desirable properties for $\Phi(x)$. 
In the diagonal gauge (\ref{diagonal}), $\theta_i(x)$ takes values in the range: 
\begin{equation}
-\pi < \theta_i(x) \le \pi,
\label{U(N) region}
\end{equation} 
and (\ref{UNphi}) can be expressed as 
\begin{equation}
\Phi^{(1)}(x)=\diag( f_1(\theta_1(x)),\cdots, f_1(\theta_N(x))  ) \qquad  \mbox{with} 
\qquad f_1(\theta)=4\tan\frac{\theta}{2}.
\end{equation}
This clearly satisfies (\ref{expansion}), and $f_1(\theta)$ 
is zero only at the origin. 
The point we could avoid the discussion above 
is that $f_1(\theta)$ is a periodic but not regular function; 
it diverges at the boundary of the region (\ref{U(N) region}). 
Another important point we should notice is that the interactions in (\ref{UNphi}) are local, 
similarly to the case of the admissibility condition (\ref{admissibility}) 
discussed in~\cite{Sugino:2004qd,Luscher:1998du}. 
This guarantees a local field theory to be obtained from the lattice model in the continuum limit. 
Actually, (\ref{UNphi}) connects link variables separated by at most a few lattice sites.%
~\footnote{ 
Note that $(U_{12}(x)-U_{21}(x))^{-1}$ is local on the lattice, 
since the inverse is taken with respect not to lattice sites but to internal gauge indices.} 

Let us next check the absence of fermion doublers. 
By plugging $U_\mu(x)=e^{iaA_\mu(x)}$ into a part of the action concerning fermion kinetic terms, i.e. 
the last line of the r.h.s. in (\ref{Slat_N=2}) with (\ref{UNphi}) used for $\Phi(x)$, 
we read off the kinetic terms of the fermions in the limit $U_\mu(x)\to 1$. 
Then, some care is needed in taking this limit for 
\begin{align}
Q\Phi^{(1)}(x)  = & -2iQ(U_{12}(x)+ U_{21}(x)) \,(U_{12}(x)-U_{21}(x))^{-1} \nn \\
&  -2i (2-U_{12}(x)-U_{21}(x))(U_{12}(x)-U_{21}(x))^{-1} \nn \\
& \quad \times Q(U_{12}(x)-U_{21}(x)) \,(U_{12}(x)-U_{21}(x))^{-1}
  + (\mbox{h. c.}). 
\label{QPhi}
\end{align}
We have 
\begin{equation}
\Delta_1\psi_2(x)-\Delta_2\psi_1(x)+F_{12}(x) (\Delta_1\psi_2(x) -\Delta_2\psi_1(x)) F_{12}(x)^{-1} 
+ {\cal O}(a) 
\end{equation}
from the first term in the r.h.s. of (\ref{QPhi}), and 
\begin{equation}
-F_{12}(x) (\Delta_1\psi_2(x) -\Delta_2\psi_1(x)) F_{12}(x)^{-1} 
+ {\cal O}(a) 
\end{equation} 
from the second term, 
where $\Delta_\mu$ is the forward difference given by 
$\Delta_\mu \varphi(x) \equiv \varphi(x+\hat{\mu})-\varphi(x)$. 
Consequently, terms containing $F_{12}(x)^{-1}$ cancel with each other, and the regular limit 
\begin{equation}
\left.Q\Phi^{(1)}(x) \right|_{U_\mu\to 1}= 2(\Delta_1\psi_2(x)-\Delta_2\psi_1(x)) 
\end{equation}
is obtained. 
By using the backward difference $\Delta_\mu^*$ 
($\Delta_\mu^* \varphi(x)\equiv \varphi(x) - \varphi(x-\hat{\mu})$), 
the fermion kinetic terms are expressed as 
\begin{align}
S_F^{\rm kin}=\frac{1}{2g_0^2}\sum_x \Tr \Bigl[ 
-\Psi(x)^T \frac{1}{2} \gamma_\mu \left(\Delta_\mu + \Delta_\mu^* \right) \Psi(x) 
-\Psi(x)^T \frac{1}{2} P_\mu \Delta_\mu\Delta_\mu^* \Psi(x) 
\Bigr], 
\label{fermion_kin}
\end{align}
where $\Psi(x)=\left( \psi_1(x), \psi_2(x), \chi(x), \frac{1}{2}\eta(x) \right)^T$, and matrices 
\begin{align}
\gamma_1&=-i \sigma_1\otimes\sigma_1, \quad 
\gamma_2=i\sigma_1\otimes\sigma_3, \quad
P_1=\sigma_1\otimes\sigma_2, \quad
P_2=\sigma_2\otimes {\bf 1}_2 
\end{align}
satisfy the algebra:  
\begin{equation}
\left\{ \gamma_\mu, \gamma_\nu \right\}=-2\delta_{\mu\nu}, \quad
\left\{ P_\mu, P_\nu \right\}=2\delta_{\mu\nu}, \quad
\left\{ \gamma_\mu, P_\nu \right\}=0. 
\label{alg_gam_P}
\end{equation}
This is the same kinetic action obtained in \cite{Sugino:2003yb}, 
and thus no fermion doubler appears.%
~\footnote{The second term of (\ref{fermion_kin}) can be regarded as a kind of the Wilson term twisted 
by the matrices $P_\mu$. 
Lattice fermion actions with such twisted terms are discussed 
in~\cite{Misumi:2013maa}.}

Notice that the lattice action has symmetries under 
\begin{itemize}
\item
lattice translation 
\item
gauge transformation
\item
$Q$-supersymmetry transformation (\ref{Q_SUSY}) 
\item
global $U(1)_R$ rotation: 
\begin{eqnarray}
& & U_\mu(x)\to U_\mu(x), \qquad \psi_\mu(x)\to e^{i\alpha}\,\psi_\mu(x), \nn \\
& & \phi(x) \to e^{2i\alpha}\,\phi(x), \qquad \bar{\phi}(x) \to e^{-2i\alpha}\,\bar{\phi}(x), \nn \\
& & H(x) \to H(x), \qquad \chi(x)\to e^{-i\alpha}\,\chi(x), \qquad \eta(x)\to e^{-i\alpha}\,\eta(x)
\end{eqnarray}
\item
reflection: $x\equiv (x_1, x_2)\to \tilde{x}\equiv (x_2,x_1)$ with 
\begin{eqnarray}
(U_1(x), U_2(x)) & \to & (U_2(\tilde{x}), U_1(\tilde{x})), \nn \\
(\psi_1(x), \psi_2(x)) & \to & (\psi_2(\tilde{x}), \psi_1(\tilde{x})), \nn \\
(H(x), \chi(x)) & \to & (-H(\tilde{x}), -\chi(\tilde{x})), \nn \\
(\phi(x), \bar{\phi}(x), \eta(x)) & \to & (\phi(\tilde{x}), \bar{\phi}(\tilde{x}), \eta(\tilde{x}))
\end{eqnarray}
\end{itemize} 
which are the same as the symmetries discussed for renormalization in~\cite{Sugino:2004qd}.%
~\footnote{Path-integral measures are also invariant under these transformations~\cite{Sugino:2006uf}.} 
Hence we can repeat the renormalization argument in~\cite{Sugino:2004qd}, and it is shown 
that no fine-tuning is required in taking the continuum limit to all orders in the perturbation theory. 
We conclude that the choice (\ref{UNphi}) 
provides a nonperturbative definition of the two-dimensional 
$U(N)$ ${\cal N}=(2,2)$ SYM theory.

\subsection{$SU(N)$ gauge group}
For the gauge group $SU(N)$, (\ref{SG_SUN}) means that the vacua for the gauge fields are determined by  
\begin{equation}
 \Phi(x)-\left(\frac{1}{N} \Tr\left(\Phi(x)\right)\right) {\bf 1}_N = 0.
 \label{vacuum eq}
\end{equation}
Unfortunately, 
it turns out that the expression (\ref{UNphi}) for $\Phi(x)$ cannot be applied here 
except for the $SU(2)$ case. 
In fact, plugging (\ref{diagonal}) with 
\begin{equation}
-\pi < \theta_i(x) \le \pi \ \ (i=1,\cdots,N-1), \quad 
\theta_N(x) = -\theta_1(x) -\cdots - \theta_{N-1}(x)  
\end{equation}
into the equation (\ref{vacuum eq}), 
we find 
\begin{equation}
U_{12}(x) = \exp\left( \frac{2\pi n i}{N} \right) {\bf 1}_N \qquad (n=0,\cdots,N-1)
\label{center solution}
\end{equation}
which are nothing but the $\Z_N$ center of $SU(N)$. 
There are still $N$ degenerate vacua for each plaquette. 

Here, we should note that in the $SU(2)$ case the degeneracy is harmless and 
(\ref{UNphi}) remains valid. 
Since the minimum at $\theta=\pi$ coincides with the singular 
point of the function $\tan \frac{\theta}{2}$ and the gauge field action around 
this solution diverges, it is decoupled from the theory.  

Motivated by this observation,  
in order to avoid the degeneracy for general $N$, we propose the following expression:  
\begin{align}
\Phi^{(M)}(x)&=\frac{2i}{M}\biggl(
\left( 2 - U_{12}(x)^{M} - U_{21}(x)^{M} \right)
\left( U_{12}(x)^{M} - U_{21}(x)^{M} \right)^{-1} \nn \\
&\hspace{10mm} +\left( U_{12}(x)^{M} - U_{21}(x)^{M} \right)^{-1}
 \left( 2 - U_{12}(x)^{M} - U_{21}(x)^{M} \right) \biggl)\nn \\
 &\equiv 
 \frac{4}{M}
\frac{ 2 - U_{12}(x)^{M} - U_{21}(x)^{M} }
{ -i \left( U_{12}(x)^{M} - U_{21}(x)^{M} \right) } 
\label{SUNphi}
\end{align}
with $M=1,2,\cdots$. 
This is applicable for any $N$ satisfying $N\leq 2M$. 
In the diagonal gauge (\ref{diagonal}), 
\begin{equation}
\Phi^{(M)}(x)=\diag( f_M(\theta_1(x)),\cdots, f_M(\theta_N(x))  ) \qquad \mbox{with} \qquad  
f_M(\theta)=\frac{4}{M}\tan\frac{M\theta}{2}.
\end{equation}
There are potential walls of infinite height 
at $\theta_i(x)=\pm\frac{\pi}{M}, \pm \frac{3\pi}{M}, \cdots$ in the gauge action (\ref{SG_SUN}) 
with (\ref{SUNphi}) used for $\Phi(x)$: 
\begin{align}
&S_G^{SU(N)} = \frac{1}{8g_0^2} \sum_{x} \Tr 
\left( \left\{\Phi^{(M)}(x)-\left(\frac{1}{N}\Tr \Phi^{(M)}(x)\right) {\bf 1}_N\right\}^2 \right) 
= \frac{2}{M^2N^2g_0^2}\sum_x\cL_G^{SU(N)}(x), \nn \\
&\cL_G^{SU(N)} (x)  \equiv \left\{(N-1)\tan\left(\frac{M\theta_1(x)}{2}\right)-
\sum_{i=2}^{N}\tan\left(\frac{M\theta_i(x)}{2}\right)\right\}^2 \nn \\
 &  \hspace{24mm} + (\mbox{cyclic permutations of }\theta_1(x), \cdots, \theta_N(x)). 
\label{gauge action_SUN} 
\end{align}
Let us consider the interval $(-\frac{\pi}{M}, \frac{\pi}{M})$ between the two walls nearest from the origin 
for each $\theta_i(x)$. 
Then, the point $(\theta_1(x),\cdots, \theta_N(x))$ can move inside the $N$-dimensional hypercube 
\begin{equation}
(\theta_1(x),\cdots, \theta_N(x))\in \left(-\frac{\pi}{M}, \frac{\pi}{M}\right)^N
\label{Nd_cube}
\end{equation}
satisfying the unimodular condition 
\begin{equation}
\theta_1(x) + \cdots +\theta_N(x)=0. 
\label{unimodular}
\end{equation}
It is easy to see that the gauge action (\ref{gauge action_SUN}) has a unique minimum at the origin 
in the region defined by (\ref{Nd_cube}) and (\ref{unimodular}). 
The region does not include any nontrivial $\Z_N$ vacuum 
because of $\frac{2\pi}{N}\geq \frac{\pi}{M}$ for $N\leq 2M$. 
In appendix~\ref{app:wall}, we show that whenever the point 
$(\theta_1(x),\cdots, \theta_N(x))$ approaches the boundary of the region, 
(\ref{gauge action_SUN}) diverges as the inverse square of distance from the boundary. 
Although there are several vacua outside the region we are considering, the growth of the potential 
near the boundary implies that the tunneling probability 
from the trivial vacuum to any other vacuum is zero. 
Namely, the trivial vacuum is effectively singled out as long as initial field configurations 
(in a numerical simulation) are around $U_{12}(x)={\bf 1}_N$ for ${}^\forall x$. 
Similarly to the $U(N)$ case (\ref{UNphi}), the interactions in (\ref{SUNphi}) are local on the lattice. 

Furthermore, it is straightforward to show 
\begin{equation}
\left.Q\Phi^{(M)}(x) \right|_{U_\mu\to 1}= 2(\Delta_1\psi_2(x)-\Delta_2\psi_1(x)) 
\label{QPhiM}
\end{equation}
by noting 
\begin{eqnarray}
U_{12}(x)^M+U_{21}(x)^M & = & 2-a^4 M^2 F_{12}(x)^2 +\cO(a^5), \nn \\
U_{12}(x)^M-U_{21}(x)^M & = & i2a^2 M F_{12}(x) +\cO(a^3). 
\end{eqnarray}
We see 
the absence of fermion doublers by 
repeating the same discussion as in the $U(N)$ case. 
The fact that the lattice action enjoys the same symmetries as mentioned in the $U(N)$ case is sufficient to 
obtain the desired continuum theory without fine-tuning at the level of perturbative expansions 
to all orders~\cite{Sugino:2004qd}. 
Therefore, we can conclude that the choice (\ref{SUNphi})
provides a nonperturbative definition of the two-dimensional 
$SU(N)$ ${\cal N}=(2, 2)$ SYM theory.


Before closing this subsection, it should be noted that (\ref{SUNphi}) with general $M$ 
can also be applied to the $U(N)$ case.%
~\footnote{(\ref{UNphi}) is included in (\ref{SUNphi}) as a special case of $M=1$.} 
There is a unique vacuum at the origin in the region (\ref{Nd_cube}). 
(Note that (\ref{unimodular}) is not imposed in the $U(N)$ case.) 
Since the gauge action (\ref{SG_UN}) with (\ref{SUNphi}) used for $\Phi(x)$ in the diagonal gauge 
\begin{equation}
S_G^{U(N)} = \frac{1}{8g_0^2}\sum_x\Tr \left(\Phi^{(M)}(x)^2\right) 
= \frac{2}{M^2g_0^2}\sum_x\sum_{i=1}^N \tan^2\left(\frac{M\theta_i(x)}{2}\right)
\end{equation}
infinitely grows near the boundary of the $N$-dimensional hypercube, 
configurations are always confined inside the hypercube once we start with initial configurations 
satisfying (\ref{Nd_cube}).  

\subsection{Convenient expressions for numerical simulation}
 
Although the expressions (\ref{UNphi}) 
and (\ref{SUNphi}) are well-defined from the mathematical point of view, 
they would not be appropriate for numerical simulations. 
In fact, since importance sampling is expected to mainly pick up configurations 
near the vacuum $U_{12}(x)={\bf 1}_N$, 
we would encounter the loss of significance in both of the numerator 
and the denominator due to $U_{12}(x) \sim U_{21}(x) \sim {\bf 1}_N$. 
In addition, 
it would take much machine time to compute the inverse of $U_{12}(x)-U_{21}(x)$ 
which becomes almost zero-matrix. 
For actual numerical simulations, it is better 
to rewrite them in a form which is apparently regular at the origin. 

In the $SU(N)$ gauge group, (\ref{SUNphi}) with $M$ even can be recast as 
\begin{align}
\Phi^{(2m)}(x)&=\frac{-i}{m}\Bigl((U_{12}(x)^m-U_{21}(x)^m)(U_{12}(x)^m+U_{21}(x)^m)^{-1} \nn \\
& \hspace{10mm} +(U_{12}(x)^m+U_{21}(x)^m)^{-1}(U_{12}(x)^m-U_{21}(x)^m)\Bigr) \nn \\
& \equiv \frac{2}{m}
\frac{ -i \left( U_{12}(x)^m - U_{21}(x)^m \right)}
{U_{12}(x)^{m} + U_{21}(x)^{m}  } 
\label{final phi}
\end{align}
with $m=1,2,\cdots$. 
This is apparently regular at $U_{12}(x)={\bf 1}_N$ and applicable to $SU(N)$ with $N\leq 4m$. 
Interestingly, (\ref{final phi}) also does the job in the $U(N)$ case for any $m$.%
~\footnote{In practice, $m=1$ will be convenient.} 
Upon using (\ref{final phi}) for numerical simulations, 
initial configurations should be chosen in the range 
$\theta_i(x)\in \left(-\frac{\pi}{2m}, \frac{\pi}{2m}\right)$ for ${}^\forall i, x$ 
in the diagonal gauge~(\ref{diagonal}).    
Then, the expression (\ref{final phi}) will give a convenient numerical means for both of the gauge groups 
$U(N)$ and $SU(N)$. 

We here add another comment on an actual numerical simulation. 
In a hybrid Monte Carlo simulation on a computer, 
a field configuration develops along the Monte Carlo time not continuously but stepwise 
by an appropriate molecular dynamics. 
Due to the effect of a finite time step, even if there are potential walls of infinite height, the 
configuration might ``jump'' over the wall from the physical vacuum to an unphysical vacuum.
Although such a phenomenon is harder to occur as the time step becomes finer, 
it is safe to check if the configuration always stays around the physical vacuum 
by measuring the distance of the link variables from the physical vacuum (the unit matrix) 
during the simulation. 
When the ``jump'' is detected, we are to reject the corresponding configuration by hand.
In case that the ``jump'' occurs frequently, the result of the simulation is no longer reliable 
because the detailed balance condition will be seriously broken by the rejection. 
In a simulation, therefore, it would be important to tune the time step of the molecular dynamics 
keeping the configuration staying inside the walls.

\section{2d $\cN=(4,4)$ lattice SYM theory without the admissibility condition}
\label{sec:N=4_woAD}
We can repeat almost all discussions in section~\ref{sec:N=2_woAD} in two-dimensional $\cN=(4,4)$ lattice 
SYM theory for the gauge groups $U(N)$ and $SU(N)$. 
The only differences are arguments for the absence of fermion doublers and renormalization. 

\paragraph{Absence of fermion doublers} 
Similarly to (\ref{QPhiM}) in the $\cN=(2,2)$ case, we obtain 
\begin{equation}
\left.Q_\pm \Phi^{(M)}(x)\right|_{U_\mu\to 1}= 2(\Delta_1\psi_{\pm 2}(x)-\Delta_2\psi_{\pm 1}(x)), 
\end{equation}
which leads to the fermion kinetic terms of the form (\ref{fermion_kin}) with 
\[\Psi(x)=\left(\psi_{+1}(x), \psi_{+2}(x), \chi_+(x), \frac12\eta_+(x),\psi_{-1}(x), \psi_{-2}(x), \chi_-(x), \frac12\eta_-(x)\right)^T
\]
and   
\begin{eqnarray}
 \gamma_1= 
\begin{pmatrix}  &  &  & -i\sigma_1 \\  &  & -\sigma_2 &  \\  & \sigma_2 &  &  \\ -i\sigma_1 &  &  &  \end{pmatrix}, & & 
 \gamma_2= 
\begin{pmatrix}  &  &  & i\sigma_3 \\  &  & -i{\bf 1}_2 &  \\  & -{\bf 1}_2 &  &  \\ i\sigma_3 &  &  &  \end{pmatrix}, \nn \\
P_1 = \begin{pmatrix}  &  &  & \sigma_2 \\  &  & i\sigma_1 &  \\  & -i\sigma_1 &  &  \\ \sigma_2 &  &  &  \end{pmatrix}, & & 
P_2 = \begin{pmatrix}  &  &  & -i{\bf 1}_2 \\  &  & -i\sigma_3 &  \\  & i\sigma_3 &  &  \\ i{\bf 1}_2 &  &  &  \end{pmatrix}. 
\end{eqnarray} 
This is valid for both of the gauge groups $U(N)$ and $SU(N)$. 
The matrices have the same form as discussed in~\cite{Sugino:2004qd} and satisfy (\ref{alg_gam_P}), 
which guarantees the absence of fermion doublers. 

\paragraph{Renormalization} 
The lattice theories with the gauge groups $U(N)$ and $SU(N)$ have symmetries under 
\begin{itemize}
\item
lattice translation 
\item
gauge transformation
\item
$Q_\pm$-supersymmetry transformations (\ref{Qpm_SUSY}) 
\item
global $SU(2)_R$ rotation generated by  
\begin{eqnarray}
J_{\pm\pm} & = & \sum_{x, \alpha}\left[
\sum_{\mu=1}^2 \psi_{\pm\mu}^\alpha(x)\frac{\partial}{\partial\psi_{\mp\mu}^\alpha(x)} 
+\chi_\pm^\alpha(x)\frac{\partial}{\partial\chi_\mp^\alpha(x)} 
-\eta_\pm^\alpha(x)\frac{\partial}{\partial\eta_\mp^\alpha(x)}\right. \nn \\
& & \hspace{9mm} \left. \pm 2\phi_\pm^\alpha(x)\frac{\partial}{\partial C^\alpha(x)} 
\mp C^\alpha(x)\frac{\partial}{\partial\phi_\mp^\alpha(x)}\right], \nn \\
J_{0} & = & \sum_{x, \alpha}\left[
\sum_{\mu=1}^2 \left(\psi_{+\mu}^\alpha(x)\frac{\partial}{\partial\psi_{+\mu}^\alpha(x)} 
-\psi_{-\mu}^\alpha(x)\frac{\partial}{\partial\psi_{-\mu}^\alpha(x)} \right) 
+\chi_+^\alpha(x)\frac{\partial}{\partial\chi_+^\alpha(x)}  \right. \nn \\
 & & \hspace{9mm}
-\chi_-^\alpha(x)\frac{\partial}{\partial\chi_-^\alpha(x)} 
+\eta_+^\alpha(x)\frac{\partial}{\partial\eta_+^\alpha(x)}
-\eta_-^\alpha(x)\frac{\partial}{\partial\eta_-^\alpha(x)} \nn \\
& &\hspace{9mm} \left. +2\phi_+^\alpha(x)\frac{\partial}{\partial \phi_+^\alpha(x)} 
-2\phi_-^\alpha(x)\frac{\partial}{\partial\phi_-^\alpha(x)}\right]   
\end{eqnarray}
($\alpha$ labels a basis of gauge group generators) which satisfy
\begin{equation}
[J_0, J_{\pm\pm}]= \pm 2J_{\pm\pm}, \qquad [J_{++}, J_{--}]=J_0 
\end{equation} 
\item 
$Q_+\leftrightarrow Q_-$ with 
\begin{eqnarray}
 & & \phi_{\pm}\to -\phi_\mp, \qquad B \to -B, \qquad \tilde{H}_\mu \to -\tilde{H}_\mu, \nn \\
 & & \chi_\pm \to -\chi_\mp, \qquad \psi_{\pm\mu}\to \psi_{\mp\mu}, \qquad \eta_\pm \to \eta_\mp 
\end{eqnarray}
\item
reflection: $x\equiv (x_1, x_2)\to \tilde{x}\equiv (x_2,x_1)$ with 
\begin{eqnarray}
(U_1(x), U_2(x)) & \to & (U_2(\tilde{x}), U_1(\tilde{x})), \nn \\
(\psi_{\pm 1}(x), \psi_{\pm 2}(x)) & \to & (\psi_{\pm 2}(\tilde{x}), \psi_{\pm 1}(\tilde{x})), \nn \\
(\tilde{H}_1(x), \tilde{H}_2(x)) & \to & (\tilde{H}_2(\tilde{x}), \tilde{H}_1(\tilde{x})), \nn \\
(H(x), B(x), \chi_\pm(x)) & \to & (-H(\tilde{x}), -B(\tilde{x}), -\chi_\pm(\tilde{x})), \nn \\
(\phi_\pm(x), C(x), \eta_\pm(x)) & \to & (\phi_\pm(\tilde{x}), C(\tilde{x}), \eta_\pm(\tilde{x})).
\end{eqnarray}
\end{itemize} 
As discussed in~\cite{Sugino:2004qd}, these symmetries are sufficient to show that the lattice theories 
do not need fine-tuning in taking the continuum limit to all orders in perturbative expansions. 

As a conclusion, the $\cN=(4,4)$ lattice model (\ref{Slat_N=4}) 
with (\ref{SUNphi}) used for $\Phi(x)$ 
nonperturbatively defines two-dimensional $\cN=(4,4)$ 
SYM theories with gauge groups $U(N)$ and $SU(N)$. 
In actual numerical simulations for the $\cN=(4,4)$ SYM theories, the lattice action with (\ref{final phi}) 
will be convenient for both of the gauge groups $U(N)$ and $SU(N)$.

\section{Summary and discussion}
\label{sec:summary}
In this paper, we have modified the lattice formulation 
of two-dimensional ${\cal N}=(2,2)$ and $(4,4)$ SYM theories 
discussed in~\cite{Sugino:2003yb,Sugino:2004qd} 
in such a way that lattice actions resolve vacuum degeneracy for gauge fields 
without imposing admissibility conditions for both of the gauge groups $U(N)$ and $SU(N)$. 
The modification yields simpler lattice actions, 
which will make numerical simulations more feasible. 

This method will also be useful to other two-dimensional supersymmetric lattice gauge 
theories with unitary link variables constructed so far -- for example, 
$\cN=(8,8)$ SYM theory with mass deformations~\cite{Hanada:2010kt} 
and $\cN=(2,2)$ supersymmetric QCD~\cite{Sugino:2008yp,Kikukawa:2008xw}. 
Since the former case is expected to give a nonperturbative construction of 
four-dimensional $\cN=4$ SYM theory~\cite{Hanada:2010kt}, it would lead to some simplification for 
the construction of the four-dimensional theory. 
For the latter case, especially the model constructed in~\cite{Kikukawa:2008xw}, 
chiral flavor symmetry of matter supermultiplets is preserved on the lattice 
by using the Ginsparg-Wilson formulation~\cite{Luscher:1998du,Ginsparg:1981bj}.   
The admissibility condition is used there for resolving the vacuum degeneracy for gauge fields 
as well as for locality of the overlap Dirac 
operator~\cite{Neuberger:1997fp,Hernandez:1998et}. 
Interestingly, it suggests that our method can also be applied to formulate chiral gauge theories 
without imposing admissibility conditions.


\section*{Acknowledgments}
The authors would like to thank 
T.~Misumi and
K.~Ohta 
for useful discussion. 
The work of S.~M. is supported in part by Grant-in-Aid
for Young Scientists (B), 23740197
and Keio Gijuku Academic Development Funds. 
The work of F.~S. is supported in part by Grant-in-Aid
for Scientific Research (C), 25400289.

\appendix
\section{Potential walls of the gauge action for $SU(N)$ gauge group}
\label{app:wall}
\setcounter{equation}{0}
In this appendix, we show that 
near the boundary of the region defined by (\ref{Nd_cube}) and (\ref{unimodular}) 
the gauge action (\ref{gauge action_SUN}) increases as 
the inverse square of distance from the boundary. 

Let us consider $\cL_G^{SU(N)}(x)$ in (\ref{gauge action_SUN}) for an arbitrary fixed $x$. 
In $N=2$ case, $|\theta_1(x)|=\frac{\pi}{M}-\epsilon(x)$ ($0<\epsilon(x)\ll 1$) 
and $\theta_2(x)=-\theta_1(x)$ near the boundary. 
Then, $\cL_G^{SU(2)}(x)$ behaves as 
\begin{equation}
\cL_G^{SU(2)}(x) = 8\tan^2\left(\frac{M\theta_1(x)}{2}\right)=8\cot^2\left(\frac{M\epsilon(x)}{2}\right)
= \frac{32}{M^2}\frac{1}{\epsilon(x)^2}+ (\mbox{finite}), 
\end{equation} 
which diverges as the inverse square of the distance from the boundary $\epsilon(x)$. 

For $N\geq 3$, when the point $(\theta_1(x),\cdots, \theta_N(x))$ approaches the boundary, 
all the possibilities are exhausted by the following cases (I) and (II): 
\\
\noindent
(I) Among the coordinates $\{\theta_1(x),\cdots, \theta_N(x)\}$, 
some coordinate (say $\theta_k(x)$) approaches $\frac{\pi}{M}$ or $-\frac{\pi}{M}$, while 
some other coordinate ($\theta_\ell(x)$) does not. \\
\noindent
(II) Every coordinate approaches $\frac{\pi}{M}$ or $-\frac{\pi}{M}$. 

By introducing an $N$-dimensional real vector $\vec{A}(x)=(A_1(x), \cdots, A_N(x))^T$ with 
\begin{equation}
A_i(x) \equiv (N-1)\tan \left(\frac{M\theta_i(x)}{2}\right)
-\sum_{j(\neq i)}\tan\left(\frac{M\theta_j(x)}{2}\right) \qquad (i=1,\cdots, N)
\end{equation}
and an arbitrary $N$-dimensional unit vector $\vec{n}=(n_1,\cdots, n_N)^T$, 
we put a lower bound to $\cL_G^{SU(N)}(x)$: 
\begin{eqnarray}
 \cL_G^{SU(N)}(x) & = & |\vec{A}(x)|^2 = |\vec{A}(x)|^2 |\vec{n}|^2 \geq (\vec{A}(x)\cdot \vec{n})^2 \nn \\
 & = & \left[\sum_{i=1}^N \left\{n_i(N-1)-\sum_{j(\neq i)}n_j\right\} \tan\left(\frac{M\theta_i(x)}{2}\right)\right]^2. 
\label{cL_bound} 
\end{eqnarray}

\paragraph{Case (I)} 
$\theta_k(x)$ satisfies $|\theta_k(x)|=\frac{\pi}{M}-\epsilon_k(x)$ with $\epsilon_k(x) \to +0$, 
and $\theta_\ell(x)$ remains to give a finite value of $\tan\left(\frac{M\theta_\ell(x)}{2}\right)$.
Use of the bound (\ref{cL_bound}) with the choice of $\vec{n}$:  
\begin{equation}
n_k=-n_\ell=\frac{1}{\sqrt{2}}, \qquad \mbox{the other components are zero}
\label{n_choice}
\end{equation}
leads to 
\begin{eqnarray}
\cL_G^{SU(N)}(x) & \geq & \frac{N^2}{2}\left\{\tan\left(\frac{M\theta_k(x)}{2}\right)
-\tan\left(\frac{M\theta_\ell(x)}{2}\right)\right\}^2 \nn \\
& = & \frac{N^2}{2}\left\{\cot\left(\frac{M\epsilon_k(x)}{2}\right)-\mbox{(finite)}\right\}^2 
= \frac{2N^2}{M^2}\frac{1}{\epsilon_k(x)^2} +\cO\left(\epsilon_k(x)^{-1}\right). 
\label{cL_bound_I}
\end{eqnarray}
Note that the expression (\ref{gauge action_SUN}) becomes no more singular than inverse squared 
with respect to distance from the boundary. 
Together with the bound (\ref{cL_bound_I}), we can conclude that $\cL_G^{SU(N)}(x)$ diverges as $\epsilon_k(x)^{-2}$. 

\paragraph{Case (II)} 
Clearly from (\ref{unimodular}), there are two coordinates $\theta_k(x)$ and $\theta_\ell(x)$ such that 
$\theta_k(x)=\frac{\pi}{M}-\epsilon_k(x)$ ($\epsilon_k(x)\to +0$) and 
$\theta_\ell(x) = -\frac{\pi}{M}+\epsilon_{\ell}(x)$ ($\epsilon_\ell(x)\to +0$). 
Then, applying the choice (\ref{n_choice}) to the bound (\ref{cL_bound}), we find 
\begin{eqnarray}
\cL_G^{SU(N)}(x) &\geq & \frac{N^2}{2}\left\{\cot\left(\frac{M\epsilon_k(x)}{2}\right)+
\cot\left(\frac{M\epsilon_\ell(x)}{2}\right)\right\}^2 \nn \\
 & = & \frac{2N^2}{M^2}\left(\frac{1}{\epsilon_k(x)} + \frac{1}{\epsilon_\ell(x)}\right)^2 +(\mbox{finite}). 
\label{cL_bound_II}
\end{eqnarray}
Similarly to the previous case, it is shown that $\cL_G^{SU(N)}(x)$ diverges 
as the inverse square of distance from the boundary.

%

\providecommand{\href}[2]{#2}\begingroup\raggedright\endgroup

\end{document}